# Continuous doping of a cuprate surface: new insights from *in-situ* ARPES

Y. -G. Zhong[1,2,†], J. -Y. Guan[1,2,†], X. Shi[1,2], J. Zhao[1,2], Z. -C. Rao[1,2], C. -Y. Tang[1,2], H. -J. Liu[1,2], Z. Y. Weng[3,4], Z. Q. Wang[5], G. D. Gu[6], T. Qian[1,4], Y. -J. Sun[1,7,*], and H. Ding[1,2,4,7,*]

[1] *Beijing National Laboratory for Condensed Matter Physics and Institute of Physics, Chinese Academy of Sciences, Beijing 100190, China*

[2] *School of Physics, University of Chinese Academy of Sciences, Beijing 100190, China*

[3] *Institute for Advanced Study, Tsinghua University, Beijing 100084, China*

[4] *Collaborative Innovation Center of Quantum Matter, Beijing 100190, China*

[5] *Department of Physics, Boston College, Chestnut Hill, MA 02467, USA*

[6] *Condensed Matter Physics and Materials Science Department, Brookhaven National Laboratory, Upton, NY 11973, USA*

[7] *CAS Center for Excellence in Topological Quantum Computation, University of Chinese Academy of Sciences, Beijing, 100190, China*

[†] These authors contributed to this work equally.

[*] Corresponding authors: dingh@iphy.ac.cn, yjsun@iphy.ac.cn

**The cuprate superconductors distinguish themselves from the conventional superconductors in that a small variation in the carrier doping can significantly change the superconducting transition temperature ($T_c$), giving rise to a superconducting dome where a pseudogap[1,2] emerges in the underdoped region and a Fermi liquid appears in the overdoped region. Thus a systematic study of the properties over a wide doping range is critical for understanding the superconducting mechanism. Here, we report a new technique to continuously dope the surface of $Bi_2Sr_2CaCu_2O_{8+x}$ through an ozone/vacuum annealing method. Using *in-situ* ARPES, we obtain precise quantities of energy gaps and the coherent spectral weight over a wide range of doping. We discover that the d-wave component of the quasiparticle gap is linearly proportional to the Nernst temperature that is the onset of superconducting vortices[3], strongly suggesting that the emergence of superconducting pairing is concomitant with the onset of free vortices, with direct implications for the onset of superconducting phase coherence at $T_c$ and the nature of the pseudogap phenomena.**

$Bi_2Sr_2CaCu_2O_{8+x}$ (Bi2212) single crystals have been extensively studied by angle-resolved photoemission spectroscopy (ARPES) and scanning tunneling spectroscopy (STS)[4,5], two of the major experimental techniques for probing the cuprates. However, high-quality Bi2212 crystals can be only obtained within a narrow doping range. Moreover, surface cleaving, necessary for surface techniques such as ARPES and STS, posses a serious problem for quantitative comparisons from sample to sample due to variation of surface conditions. Realizing that the doping level in this material is solely controlled by the excess oxygen concentration, we use ozone/vacuum annealing to continuously change the doping level of the surface layers, which are subsequently measured by *in-situ* ARPES (Figs. **1a-c**).

It is well established that the electronic structure of the doped $CuO_2$ planes of Bi2212 exhibits two-dimensional Fermi surface (FS) sheets whose area relative to the Brillouin Zone area is equal to $(1+x)/2$, where $x$ is the doped carrier concentration[6]. Therefore, a precise determination of FS area by high-resolution ARPES, taking extra care of the super-lattice and shadow FSs (Fig. **1c**) (details see Supplementary Information (SI)), will provide the value of carrier concentration on the surface. Figures **1d-g** illustrate an example of how we continuously change the doping level and the corresponding FS contours on one surface. Here we point out a few noticeable features of the FS evolution: 1. There is a continuous reduction of the spectral weight on the FS going from overdoped (OD) to underdoped (UD), especially around the antinodal region, which is mainly caused by increasing of the superconducting gap and emergence of the pseudogap in the UD region. 2. The bilayer splitting[7], clearly visible in the OD region, gradually vanishes in the UD region due to the incoherent c-axis tunneling between adjacent $CuO_2$ planes. 3. The shrinkage of the FS area can be clearly visualized by the synchronized movement of the main FS and two super-lattice FSs along the nodal direction (Fig. **1e**). A simple fitting to the measured FS (details see SI) yields the carrier concentration of the surface (Fig. **1f**). Remarkably, a wide doping range within a nearly full SC dome can be continuously tuned on a single Bi2212 surface (Fig. **1g**). Such a "Phase-Diagram-on-Surface" (PDS) method, not only extends the doping range beyond the conventional single crystal method, but also eliminates the uncontrolled influence to ARPES spectral weight due to different flatness of cleaved surfaces, thus enabling precise analysis and comparison of important quantities over the phase diagram of the cuprates.

One of the important quantities is the quasiparticle spectral weight (Z), which has been studied extensively by ARPES[8,9]. Despite of intensive efforts, reliable quantitative results of the doping evolution of Z and its momentum dependence are difficult to obtain due to its sensitivity to the surface condition that varies from sample to sample. The new PDS method overcomes this problem. A coherent quasiparticle peak emerges in the superconducting state (Figs. **2a, b**), especially around the antinodal region, whose spectral weight ($Z_{AN}$) is believed to relate to the superfluid density or the superconducting phase stiffness[10]. Previous ARPES studies[8,9] have indeed revealed that $Z_{AN}$ is proportional to $x$ below the optimal doping level. However, the behavior of $Z_{AN}$ on the overdoped side is controversial, with initial reports of saturating[8] or decreasing[9] behaviors. Figures **2a**, **c** reveal that $Z_{AN}$ on the antinodal FS is linearly proportional to $x$ within the entire superconducting dome. We note that a saturating behavior is observed for the coherent weight at the M point that is not on the FS (details see SI). In the meantime, the PDS method also enables a precise measurement of the nodal quasiparticle weight ($Z_N$) that is known to be highly sensitive to the surface flatness. It is clear from Figs. **2b**, **c** that the doping evolution of $Z_N$ is much milder than that of $Z_{AN}$: it stays almost constant for a wide region on both sides of the optimal doping level, and decreases in the more underdoped region, possibly due to the appearance of Coulomb gap or other incoherent processes with heavy underdoping[11].

Another important quantity that can be systematically studied by the PDS method is the quasiparticle energy gap in the one-electron spectral function, which is a manifestation of both the superconducting gap and the possible pseudogap[4,5,12]. Since sharp quasiparticle peaks can be observed in the superconducting state at low temperature, their peak position can be used as a good measure of the energy gap. The symmetrized peaks[13] remove the effect of the Fermi-Dirac function and thus give more precise values of energy gap along the FS. We summarize in Figs. **3a-f** the momentum dependence of the energy gap over a wide doping range ($0.07 \leq x \leq 0.24$). In the OD region, e.g., $x = 0.24$ and $0.21$, the momentum dependence of the energy gap follows the $\cos k_x$-$\cos k_y$ function nicely, reflecting the nature of a single d-wave pairing gap in this region. However, starting from $x = 0.18$, the energy gap extracted from the quasiparticle position deviates upward from the simple d-wave function, and the deviation increases as the doping decreases[14]. This is indicative of the opening of the pseudogap whose origin has been widely debated[15,16]. It has been pointed out that the underlying superconducting gap follows the d-wave line visibly near the nodal region and extrapolates to the antinodal region, namely it follows the gap slope ($\Delta_0$)[17,18]. We adopt this method and extrapolate the superconducting gap from the node to the antinode (details see SI) for all the doping levels (Fig. **3a-f**). If we regard the quasiparticle peak position as the superposition by quadrature of two energy gaps, as suggested previously[19,20], we can then decompose the total spectral gap ($\Delta_{tot}$) into two gaps, $\Delta_{tot} = \sqrt{(\Delta_0^2 + \Delta_{PG}^2)}$, where $\Delta_0$ and $\Delta_{PG}$ are the pairing gap and the pseudogap, respectively (Fig. **3f**). While the pairing gap, which is extrapolated from the nodal region, follows the d-wave function throughout the superconducting dome, the pseudogap opens up first at the antinodal region in the slightly OD region, and spreads toward the nodal region as the doping is reduced. We note that such a decomposition procedure is consistent with the observation of the pseudogap and Fermi arc phenomena in the normal state above $T_c$ in the UD regime.

A previous ARPES work[18] on single crystals of several doping levels suggested that this extrapolated energy gap is linearly proportional to $T_c$. However, a more precise study using our PDS method gives a qualitatively different result. Through a careful comparison of the gap at antinode ($\Delta_{AN}$) to the extrapolated gap ($\Delta_0$), we find that the value of the extrapolated d-wave pairing gap $\Delta_0$ locates systematically in between the antinodal gap $\Delta_{AN}$ and the energy gap that would scale with $T_c$ (Fig. **4a**). In the OD region, these two gaps and the $T_c$-scaled gap match very well. However, they start to diverge in the UD region, and the coupling strengths $2\Delta/k_B T_c$ obtained using these two gaps increase rapidly in the UD region (Fig. **4b**), suggesting that neither $T^*$ nor $T_c$ are the thermal correspondence of the nodal d-wave gap in the quantum electronic state at low temperature. Remarkably, if we replace $T_c$ with the Nernst temperature $T_\nu$ (ref. 3), which was generally identified as a sign of Cooper pair formation, then $2\Delta_0/k_B T_\nu$ is nearly a constant over the whole SC dome (Fig. **4b**), strongly indicating that the extrapolated gap $\Delta_0$ represents a pairing energy gap associated with the formation of Cooper pairs, with a thermal correspondence to $T_\nu$ (Fig. **4c**).

While we showed that the extrapolated nodal d-wave gap as the pairing energy scale related to $T_\nu$, the origin of the pseudogap highlighted by the spectral gap around the antinode in the UD region remains an open question. Our measurements alone cannot rule out the scenario that the pseudogap stems from preformed pairs[21]. However, the divergent behavior of $\Delta_{PG}$ and $\Delta_0$, naturally supports the notion that the pseudogap is due to competing order, such as the in the proposed scenarios of charge or bond order[22], valence bond glass[23], pair density wave[24], and loop current[25]. Recent experiments have indeed provided strong evidence for charge order in the UD region[22]. Although, we have not observed direct evidence for charge order here, there is a residual spectral weight build up below the antinodal coherent peak, forming an "antinodal foot" in the UD region (Fig. **2a** and Fig. **S5** in SI). It would be interesting in the future to study whether this antinodal foot is related to the charge order.

Our systematic measurements reveal a novel two-component coherent peak structure near the Fermi level over a wide range of doping. One has a d-wave-like gap $\Delta_0$ near the nodal regime with an almost constant coherent peak weight versus doping. But the characteristic energy $\Delta_0$ itself is shown to scale with the Nernst temperature $T_\nu$, rather than $T_c$, by a ratio $2\Delta_0/K_B T_\nu \sim 6$ over the SC dome (Fig. **4b**). Here $T_\nu$ can be interpreted as the onset of the pairing transition before pairing coherence is established, while the Nernst signal comes from the vortices of the local pairing order parameter above $T_c$. Namely the difference between $T_\nu$ and $T_c$ is due to the destruction of superconductivity by vortex fluctuations instead of vanishing $\Delta_0$, and the true SC phase coherence is established at $T_c$. The region between $T_\nu$ and $T_c$ is anomalously large compared to conventional BCS superconductors, which is possibly a reflection of the energetically favorable vortex core states in the cuprates[26]. Note that the value of the ratio ~ 6 is comparable to the ratio of $2\Delta_0/K_B T_c$ between 4.6 and 5.6 (ref.

27) in the heaviest elemental superconductors such as Hg and Ir, which are in the strong coupling regime.

In contrast, an antinodal coherent peak with its spectral weight linearly proportional to $x$ is observed to persist beyond the optimal doping. Previously, the antinodal weight has been argued to relate to the superfluid density in the SC state[8-10]. However, the superfluid density has been carefully measured in the overdoped regime of La-214 and found to decrease with the reduction of $T_c$[28]. Thus, our finding indicates that the antinodal spectral weight scales with the total carrier density $x$ doped into the Mott insulating parent state over the entire superconducting compositions. In the underdoped to optimally doped regions, they condense into the superfluid, whereas a significant portion may fail to condense in the overdoped region in sharp contrast to the BCS theory. It remains to be seen whether theories of doped Mott insulators[26,29] may account for the present ARPES observations.

**Acknowledgements:**

We thank J. -J. Li, W. -Y. Liu, R. -T. Wang, J. -Q. Lin, F. -Z. Yang and S. -F. Wu for technique assistance and thank Q. J. Chen, K. Levin, J. X. Li, M. Randeria, F. C. Zhang for useful discussions. This work at IOP is supported by the grants from the Ministry of Science and Technology of China (2016YFA0401000, 2016YFA0300600, 2015CB921300, 2015CB921000), the Natural Science Foundation of China (11574371, 11622435, 11474340), the Chinese Academy of Sciences (XDB07000000, XDPB08-1, QYZDB-SSW-SLH043), and the Beijing Municipal Science and Technology Commission (Z171100002017018). Z. Q. Wang is supported by the U.S. Department of Energy, Basic Energy Sciences Grant No. DE-FG02-99ER45747. G. D. Gu is supported by the office of BES, Division of Materials Science and Engineering, U.S. DOE, under contract Nos. de-sc0012704.

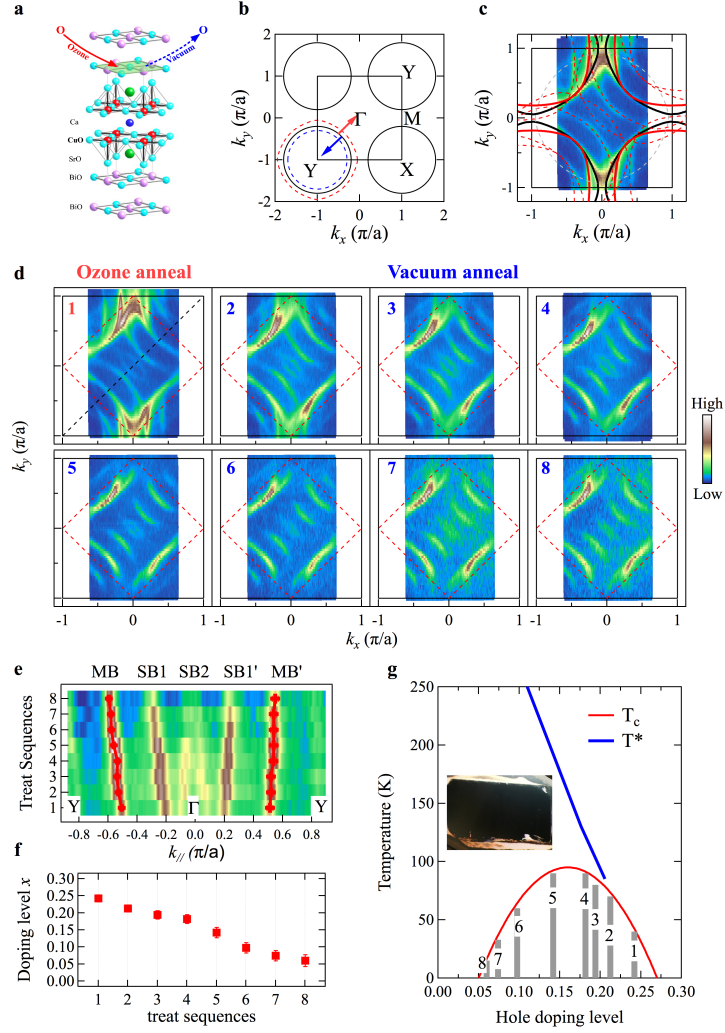

**Figure 1 | Example of tuning surface doping through PDS method. a.** Crystal structure of Bi2212, and the schematic of ozone annealing (red arrows) and vacuum annealing (blue arrows) surface treatments. **b.** The schematic FS of Bi2212 after ozone (red) and vacuum annealing (blue). **c.** The FS of the OD one with considerations of super-lattice FS (red dash line) and shadow FS (gray dash line). The bold black and red lines are the fitting results corresponding to the antibonding and bonding FS, respectively. Those FSs are acquired at 30K with integrating ± 10meV around $E_F$. **d.** The FS evolution with surface treat sequences: after an ozone annealing (panel 1), a serials of vacuum annealing step by step (panels 2-7). **e.** Image plot of the integrated intensity in the vicinity of $E_F$ along the Γ-Y direction for each panel in **d**, showing two main bands (MB, MB'), two 1$^{st}$ super-lattice bands (SB1, SB1'), and two 2$^{nd}$ super-lattice bands (SB2, SB2'). **f.** The extracted doping level of each sequence. **g.** The phase diagram of Bi2212: the red line is $T_c$ calculated with an empirical formula[30], the blue line is $T^*$ extracted from the antinodal gap closed temperature, the gray bars are the doping level for each sequence obtained in **f**, and the insert picture is the one taken on the cleaved surface of this sample.

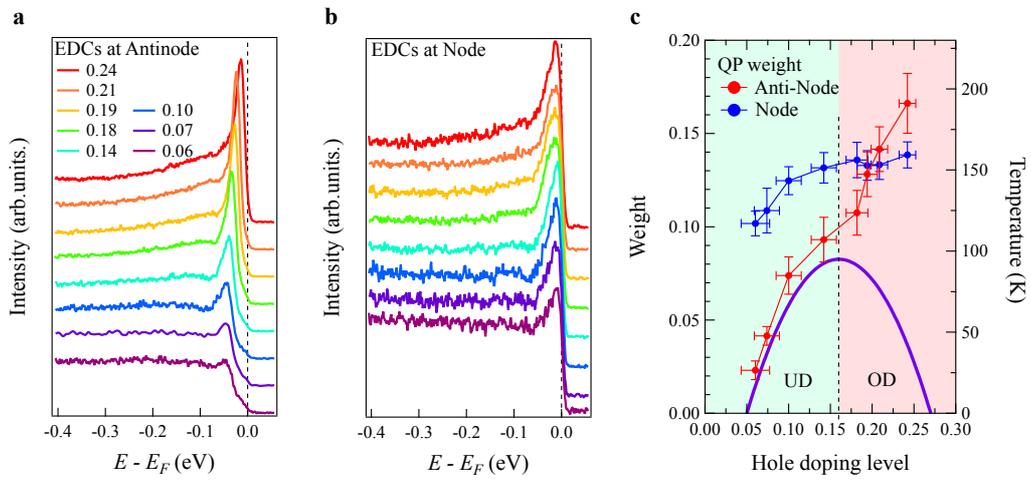

**Figure 2 | Quasiparticle weight evolution with doping. a.** Antinodal EDCs at $k_F$ evolution with doping. **b.** Nodal EDCs at $k_F$ evolution with doping measured on the same sample at 10K. **c.** The extracting quasiparticle weight of the node and the antinode. The $T_c$ line is plotted with the violet color.

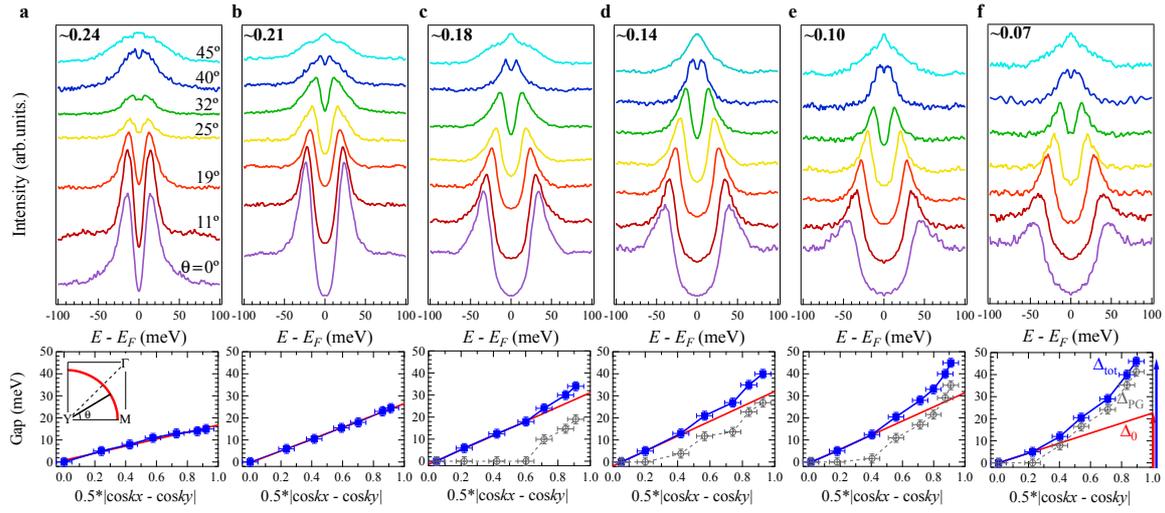

**Figure 3 | Superconducting gap along FS with different doping levels. a-f.** Upper panels display the symmetrized EDCs along the FS of six different doping levels. All data are measured on a same sample and a same temperature of ~10K. The corresponding momentum positions are indicated in a schematic diagram of FS in the inset of the lower panel of **a**. Lower panels display the spectral gap along the FS (plotted against $0.5*|\cos k_x - \cos k_y|$) for different doping levels. The blue marked one is the total spectral gap, which is determined from the peak position of symmetrized EDCs, the red line is the corresponding d-wave gap function (details see SI) for each doping, and the gray marked one is the pseudogap obtained as describe in the text.

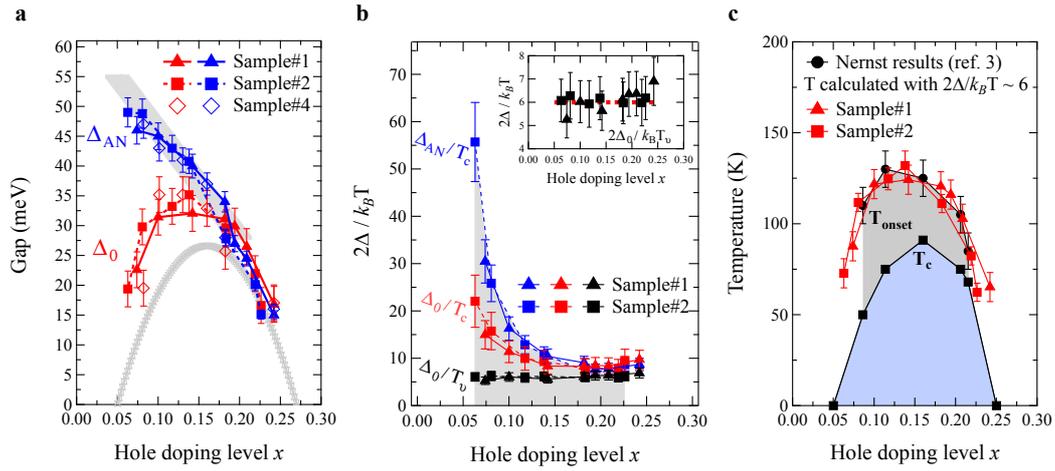

**Figure 4 | Pairing gap under antinodal gap. a.** Doping dependence of the antinode gap $\Delta_{AN}$ and d-wave gap slope $\Delta_0$. The red symbols are $\Delta_0$, the blue ones $\Delta_{AN}$, and the gray ones the $T_c$-scaled gap, with three different symbols (solid square, triangle and spaced rhombus) for three independent samples. **b.** Doping dependence of $2\Delta/k_B T_c$. The blue symbols are $2\Delta_{AN}/k_B T_c$, the red ones $2\Delta_0/k_B T_c$, the black ones $2\Delta_0/k_B T_v$. The insert is a zoom-in plot to highlight the constant ratio of $2\Delta_0/k_B T_v$. **c.** A phase diagram of Bi2212. The red symbols are the pairing temperatures obtained from the constant ratio with $\Delta_0$, the black dots the onset temperatures of the Nernst signal from ref. 3, black square ones the superconducting transition temperatures.

## Method

### Surface treatment

The optimally doped Bi2212 single crystals were grown by the floating-zone technique. Heating samples in an ultrahigh vacuum and ozone atmosphere are referred as the vacuum annealing and ozone annealing, respectively. A high-quality optimally doped single crystal was cleaved in a vacuum better than $1\times10^{-7}$ Torr, degased in a MBE chamber to ensure surface fresh, and annealed at ~470°C at the ozone atmosphere with a partial pressure about $4\times10^{-6}$ Torr for 15 minutes to obtain a highly overdoped surface. Then the doping level of the surface can be reduced by subsequent vacuum annealing processes, with the annealing temperature controlling the surface doping level. After each annealing process, the sample was transferred *in-situ* to an ARPES chamber for ARPES measurements. Freshness of the sample surface can be regenerated through each annealing process, and the sample surface can be measured over one month without noticeable contaminations.

### ARPES measurement

The *in-situ* ARPES measurements were carried in our ARPES system equipped with a Scienta R4000 analyzer and a Scienta VUV source. The He Iα resonant line (hυ= 21.218 eV) was used, and the vacuum of the ARPES chamber was better than $3\times10^{-11}$ Torr. The energy resolution was set at ~ 5meV and the angular resolution was 0.2°.

# Supplementary Materials for "Continuous doping of a cuprate surface: new insights from *in-situ* ARPES"

Y. -G. Zhong[1,2,†], J. -Y. Guan[1,2,†], X. Shi[1,2], J. Zhao[1,2], Z. -C. Rao[1,2], C. -Y. Tang[1,2], H. -J. Liu[1,2], Z. Y. Weng[3,4], Z. Q. Wang[5], G. D. Gu[6], T. Qian[1,4], Y. -J. Sun[1,7,*], and H. Ding[1,2,4,7,*]

[1] Beijing National Laboratory for Condensed Matter Physics and Institute of Physics, Chinese Academy of Sciences, Beijing 100190, China

[2] School of Physics, University of Chinese Academy of Sciences, Beijing 100190, China

[3] Institute for Advanced Study, Tsinghua University, Beijing 100084, China

[4] Collaborative Innovation Center of Quantum Matter, Beijing 100190, China

[5] Department of Physics, Boston College, Chestnut Hill, MA 02467, USA

[6] Condensed Matter Physics and Materials Science Department, Brookhaven National Laboratory, Upton, NY 11973, USA

[7] CAS Center for Excellence in Topological Quantum Computation, University of Chinese Academy of Sciences, Beijing, 100190, China

**The supplementary materials include the following contents:**

I. **Fermi surface fitting procedure**
II. **Extracting procedure of quasiparticle weight**
III. **Gap slope around the node**
IV. **Leading edge foot of the EDCs in the UD region**

## 1. Fermi surface fitting procedure

As a typical hole-doped cuprate superconductor, $Bi_2Sr_2CaCu_2O_{8+x}$ (Bi2212) has a hole-liked Fermi surface sheet around $(\pi, \pi)$ in the Brillouin Zone with the following features: 1. Supper-lattice Fermi surface (FS) sheets, which are commonly regarded as originated from diffraction of the photoelectrons off the super-lattices present in the BiO planes, behaving like the replicas of the main FS sheet shifted by $Q = \pm (0.21\pi, 0.21\pi)$ along the ΓY direction[S1,S2], as the red dash lines shown in Fig. S1a. 2. The shadow FS sheets form by folding the main FS sheets with respect to the antiferromagnetic Brillouin zone boundary due to magnetic or structural causes[S3,S4], illustrated by the gray dash line in Fig. S1a. 3. Bilayer splitting[S5], resulted from the coherent c-axis tunneling between adjacent $CuO_2$ planes, which is more pronouncing in the OD region, as illustrated in Fig. S1a.

We fit the FS using the one-band tight binding model[S6] in order to obtain the FS area. Considering the following energy dispersion function (Eq.1), and setting $E(k_x, k_y) = E_F$, we can get a function relationship between $k_x$ and $k_y$, which gives the morphology of the FS.

$$E(k_x,k_y)=\mu+\frac{t_0}{2}(\cos k_x a+\cos k_y a)+t_1 \cos k_x a \cos k_y a+\frac{t_2}{2}(\cos 2k_x a+\cos 2k_y a)\pm t_{bi}(\cos k_x a-\cos k_y a)^2 \cdots\cdots(1)$$

In our fitting, we use four free parameters, $\mu$, $t_0$, $t_1$, and $t_{bi}$. Since the $t_{bi}$ term in the energy dispersion function can be merged into $t_1$ and $t_2$ terms through decomposing, we fit the bonding FS and the anti-bonding FS independently and sum their areas to obtain the total FS area for the OD region. Since in the UD region, the antinodal intensity integrated around the Fermi energy is weak due to a large gap opening, in order to get the doping level precisely, we also fit the real FS intensity images integrated around the minimum gap energy, as illustrated in Figs. S2a-d. As shown in Fig. S1a and Fig. S2a, the fitted FS not only follows the main FS well, but also its Q-shifted FS traces the super-lattice FS nicely. We then calculate the area of the fitting FS sheet to obtain the area of the measured FS sheet, and consequently determine the doping level. We emphasize that this fit procedure is reasonable if one only cares about the FS area. The fitting FS sheets for all the treating sequence of one sample surface are summarized in Fig. S1b-c. One can clearly observe the continuous shrinkage of the FS area. To make the results more reliable, we repeat the ozone/vacuum annealing experiments on other three optimally doped samples, and obtain their surface carrier concentrations, as shown in Table I.

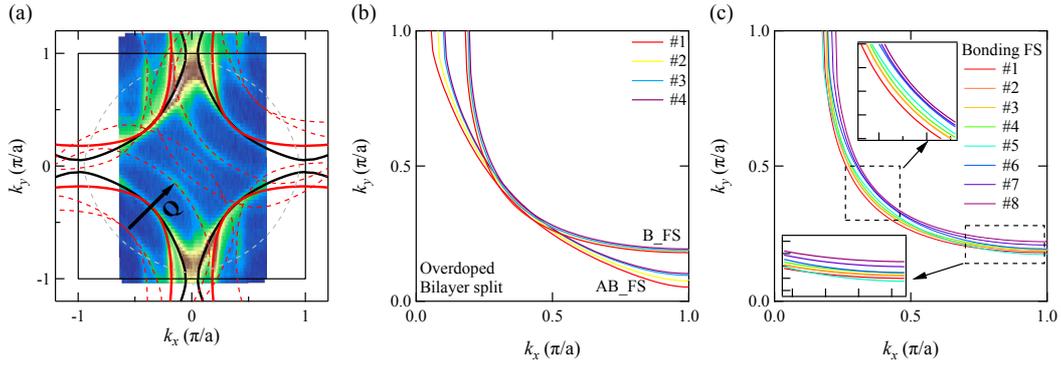

**Figure S1. Fitting results of Fermi surface.** a. The measured FS and the fitting result of an OD Bi2212. The red bold lines are the fitting curves to the bonding FS, and the black bold lines are to the antibonding FS, the red dash lines are the Q-shifted (Q = (±0.21π, ± 0.21π)) fitting curves to the super-lattice FS, and the gray dash lines are the (π, π)-folded fitting curves to the shadow FS. b. The fitting results of the ozone/vacuum annealing sequences #1 to #4 that are within the OD region, the bonding FS and antiboding FS sheets are fitted independently as described in the text. c. The comparison of the bonding FS of all treatment sequences, and two inserts are the zoom-in figures of the nodal region and the antinodal region.

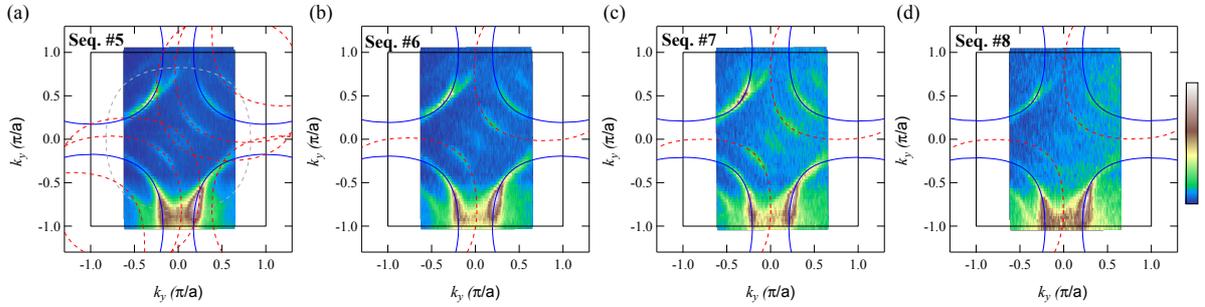

**Figure S2. The spectrum integral along the minimum gap for the UD region.** (a)-(d) are the spectral intensity images integrated around the minimum gap energy, corresponding to treat sequence #5-#7 of sample #1. The integral range is ± 5meV. The blue line (main band) and dash red line (super-lattice band) are the fitting results, matched well with the main FS and super-lattice FS. For comparison, the spectrum integral around the Fermi energy is left in the upper half part of the Brillouin Zone.

|              | Sample #1 | Sample #2* | Sample #3 | Sample #4 |
|:---:|:---:|:---:|:---:|:---:|
| orientation  | ΓM        | ΓM         | ΓM        | ΓY        |
| Seq. #1      | ~0.24     | ~0.23      | ~0.24     | ~0.24     |
| Seq. #2      | ~0.21     | ~0.22      | ~0.23     | ~0.18     |
| Seq. #3      | ~0.19     | ~0.18      | ~0.21     | ~0.15     |
| Seq. #4      | ~0.18     | ~0.14      | ~0.20†    | ~0.13     |
| Seq. #5      | ~0.14     | ~0.12      | ~0.18     | ~0.10     |
| Seq. #6      | ~0.10     | ~0.08      | ~0.11     | ~0.08     |
| Seq. #7      | ~0.07     | ~0.06      |           |           |
| Seq. #8      | ~0.05     |            |           |           |

\* Measuring focus on the antinodes.
† Just measured antinode point.

**Table I.** Summary of the calculating carrier concentration of each treating sequence for four different samples.

## 2. Extracting procedure of quasiparticle weight

Since the sharp coherent quasiparticle peak emerges below $T_c$ in Bi2212, the ARPES spectral function can be separated into the coherent and incoherent components everywhere along the FS[S7]. For an accurate extraction of the coherent peak weight, the important procedure is how to determine the incoherent part. For simplicity, here we use a linear background to represent the incoherent part under the sharp peak since the sharp peak is within a narrow energy range, as shown in Fig. S3e. In order to determine the starting energy position of the linear background, we calculate the derivative of EDCs, and then chose the energy position where the derivative is increasing suddenly as the starting energy position, as illustrated in Figs. S3e and S4d. To increase the reproducibility, we repeated this experiments on four different samples, which yield consistent results, as illustrated in Figs. S3 and S4. Sample#2 and smaple#3 are aligned to the ΓM direction, with sample#2 focusing on the antinodal weight, and sample#3 focusing on the overdoped regime. Sample#4 is aligned to GY, which makes the nodal results more accuracy.

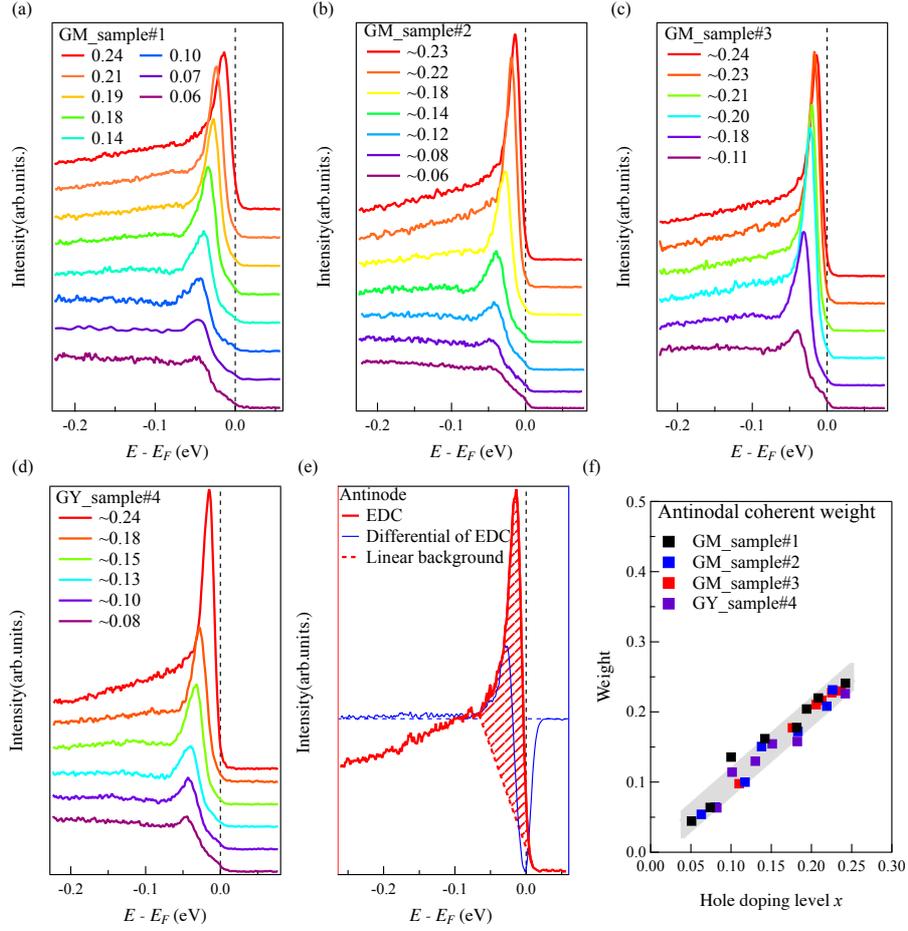

**Figure S3. Antinodal quasiparticle weight.** a-d. The antinodal EDCs of four independent ozone/vacuum annealing experiments, and their quasiparticle weight are extracted as displayed in panel f by subtracting a linear incoherent background under the sharp coherent peak as shown in panel e.

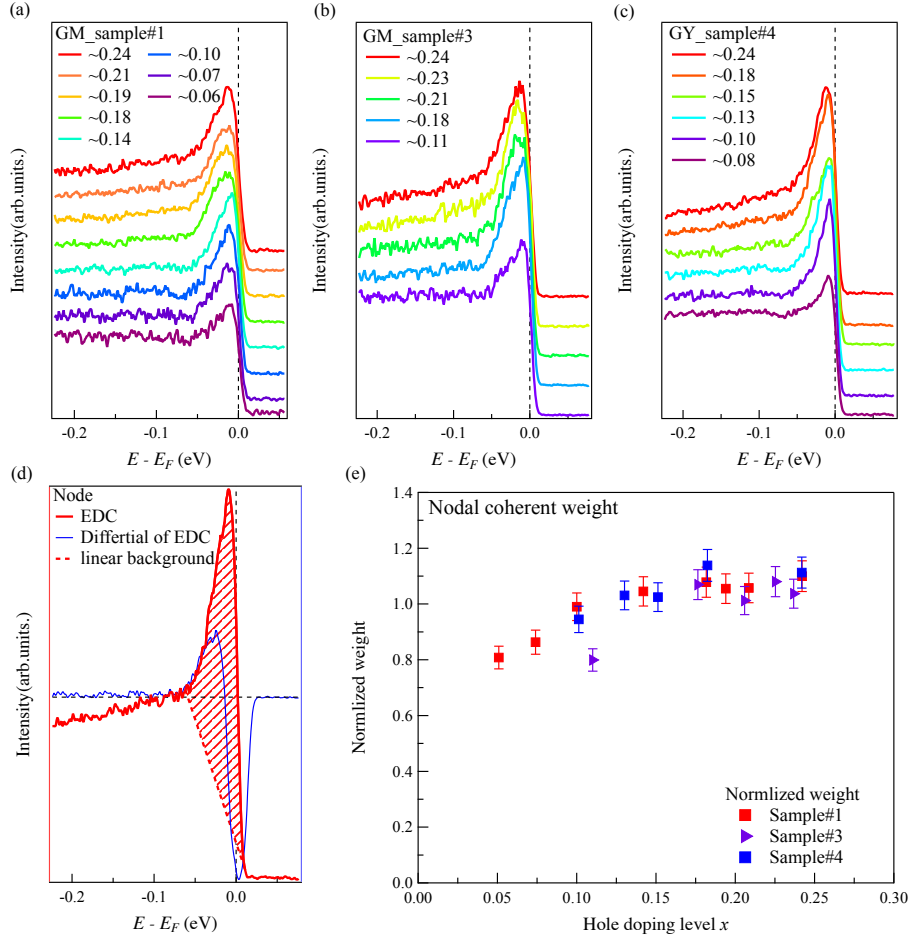

**Figure S4. Nodal quasiparticle weight.** a-c. The nodal EDCs of three ozone/vacuum annealing treating experiments, and their quasiparticle results are extracted as displayed in panel e. The extracting procedure is illustrated in panel d.

## 3. Gap slope around the node

In order to extract the d-wave gap around the nodal region, we apply the following two methods. One is to simply fit the linear portion of the gap curve with the linear d-wave gap function (Eq. 2), as shown in Fig. S5d. The second method is to fit the gap curve to a next higher-harmonic (NHH) d-wave function[S8] (Eq. 3) over the entire region, and to use its derivate at the node as the gap slope, as shown in Fig. S5d. These two methods give consistent results as shown in Figures S5e-g for different samples.

$$\Delta(kx,ky) = \Delta_0 \cos 2\theta \sim \Delta_0 |\cos k_x - \cos k_y|/2 \cdots\cdots(2)$$
$$\Delta(kx,ky) = \Delta_0 [B\cos 2\theta + (1-B)\cos 6\theta] = \Delta_0 [(4B-3)\cos 2\theta + 4(1-B)\cos^3 2\theta] \cdots\cdots(3)$$

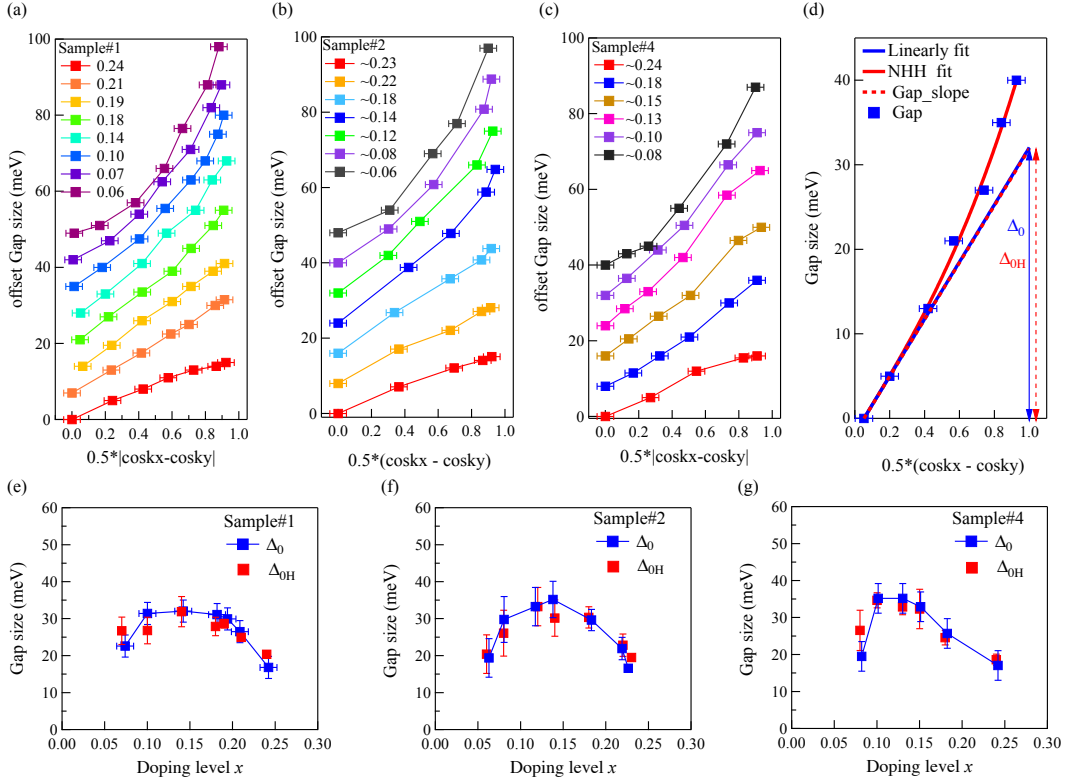

**Figure S5. Extracting procedure for the gap slope.** a-c. Gap curves as represented by quasiparticle positions along the FS plotted as functions of $cosk_x$-$cosk_y$ for samples #1, #2, #4. d. An example to extract the gap slope around the node. The blue line is a linear line representing the gap slope around the node, namely $\Delta_0$, and the red line is a fit of the gap curve to the NHH d-wave formula (Eq. 3) over the entire k region, whose smooth curve can be used to generate a derivative at the node, namely $\Delta_{0H}$. e-g. Extracted d-wave gap slopes for different samples. The blue square marks represent $\Delta_0$, and the red square marks are $\Delta_{0H}$.

## 4. Leading edge foot of the EDCs in the UD region

In the antinodal region of the underdoped sample surfaces, we consistently observed a residual spectral weight between the sharp quasiparticle peak and the Fermi energy, forming the "leading edge foot" as shown in Fig. S6a. To extract this "foot", we simulate an EDC according to Eq. 4 as illustrated in Fig. S6b, in which a Gaussian function represents the quasiparticle peak and a broad Lorentzian with an asymmetric cut-off represents the tail of the EDCs[S9].

$$I(w) = (A_{Lor} + B_{Guass}) * f_T(w) \otimes R(w) \cdots\cdots (4)$$

From the comparison between the raw EDC and the simulated EDC, we can see more clearly the "leading edge foot" as shown in Fig. S6b. Through extracting the difference of them, we obtain the spectrum of the "foot", as shown by the shadow part in Fig. S6b. The difference curves of each doping are plotted in Fig. S6c. We note that both the peak position and the area of the "foot" have a dome shape, as shown in Fig. S6d. While the origin of the "leading edge foot" is unclear, we speculate that this foot might be related to the charger order observed in the underdoped region.

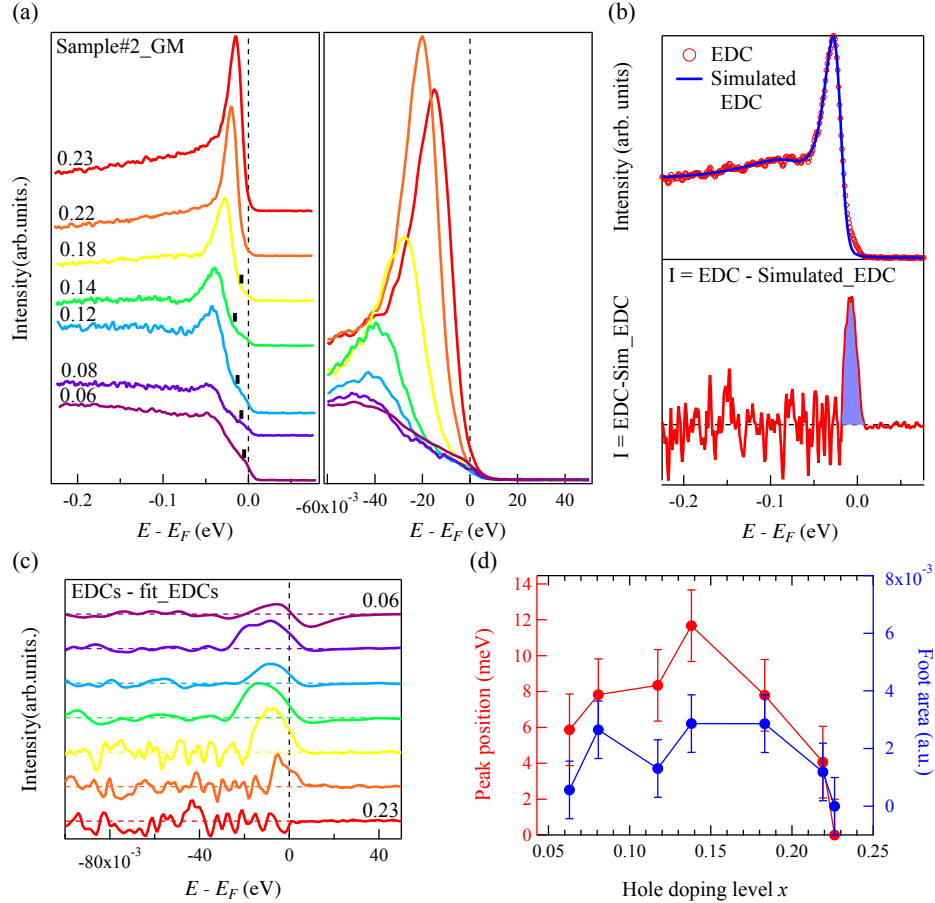

**Figure S6. Observation of a leading edge foot.** a. The left panel is the antinodal EDCs of sample #2 with different doping levels, the right one is the zoom-in plots of those EDCs without offsets.   b. The upper panel is a simulation of the EDC compared to the raw EDC, and the lower one is the difference between them.   c. The difference curves for each doping.   d. The peak positions and areas of the "foot" versus the doping level.